# HIGHER DIMENSIONAL COMMUNICATION AND S.E.T.I

Paul S. Wesson

Dept. of Physics and Astronomy, University of Waterloo, Waterloo, Ontario N2L 3G1 Canada

Abstract: In cosmologies with more than four dimensions, of the type required for unification, it is possible for signals to have velocities in excess of that of light.  Using a five-dimensional model which otherwise agrees with observations, two subjects are reviewed: (a) An exact solution of the field equations which describes a 4D spacetime with a large cosmological constant and waves travelling in ordinary 3D space with velocities exceeding lightspeed.  (b) An example where the 4D interval or proper time is modulated by the systematic variation of the scalar field associated with the fifth dimension, providing a simple signalling method.  These and related consequences of higher-dimensional cosmology have significant implications for astrophysics, and especially the Search for Extraterrestrial Intelligence.

Email:	psw.papers@yahoo.ca.

# HIGHER DIMENSIONAL COMMUNICATION AND S.E.T.I

1. Introduction

The main motivation for extending general relativity beyond space and time is to unify gravity with the interactions of particles. The hope is that higher dimensions may incorporate the symmetry groups displayed by elementary particle physics (Halpern 2004, Carr 2007). However, the fifth dimension is widely believed to engender effects on both large and small scales, and it can be constrained by matching the theory to cosmological observations. For example, the fifth dimension leads to new insight about the cosmological constant, and its influence is compatible with recent type Ia supernova data (Overduin et al, 2007). There are in fact diverse consequences of the fifth dimension, because it involves a scalar field which can affect particle masses. It has already been pointed out that it can modify the perihelion advance of Mercury (Liu and Wesson 1992), provide new candidates for dark matter (Wesson 1994) and yield a clearer picture of the big bang (Wesson and Seahra 2001). In the present account, the aim is to elucidate another new consequence of the fifth dimension, namely the possibility of signals that exceed the velocity of light.

Signals travelling at the speed of light take approximately four years to reach us from nearby star systems such as Proxima Centauri, and about sixty thousand years to span the Milky Way. The time lag caused by the finite speed of light means that if the Search for Extraterrestrial Intelligence were to detect an electromagnetic signal from creatures like us in the Andromeda galaxy, we would see them as they were about two million years ago when they resembled ape-men. Clearly, any form of communication with aliens which is to take place on a human timescale implies some kind of signal whose *effective* speed in ordinary 3D space exceeds that of light (Cosmovici et al 1997). Recently, some work on the nature of the vacuum in 5D showed the feasibility of waves that propagate with arbitrarily high velocities (Wesson 2013). It is hoped to outline the implications of these waves for astrophysics in general and S.E.T.I. in particular.

In addition to the aforementioned wave, the general 5D theory admits a scalar field which can in principle be used to transmit signals. This happens because, unlike its predecessors, modern 5D cosmology has a fifth dimension which is not compactified to an unobservably small size, but is large and manifests itself along with gravity. The 5D metric has, in general, both gravitational and scalar potentials. The path of a particle or wave through these fields is given, as in Einstein's theory, by a geodesic. However, it is now generally accepted that *null* geodesics in 5D correspond to the null paths of photons *and* the timelike paths of massive particles in 4D. This effectively redefines causality, with consequences for both interstellar communication and the evolution of the early universe.

The focus here is on astrophysics rather than gravitation, but reviews of the underlying theory are available (Halpern 2004, Carr 2007, Wesson 2008). The approach is known as Space-Time-Matter theory or Membrane theory, depending on whether the emphasis is on cosmology or particle physics. These have a common mathematical framework, which is basically an extension of general relativity from 4 to 5 dimensions. The 5D picture smoothly includes the conventional 4D one in accordance with an embedding theorem of Campbell. This embedding also ensures that the 5D theory conforms to observations at some level. But of course, before one or more extra dimensions can be accepted, it will be necessary to show some effect which is not present in 4D spacetime. Signals travelling at velocities in excess of lightspeed represent one such effect.

The vacuum waves and scalar field discussed below involve novel physics and their detection will require the development of new instrumentation. The waves do not resemble electromagnetic ones, but are more akin to the de Broglie or matter waves demonstrated in modern versions of the classic double-slit experiment (Edamatsu et al 2002, Kocsis et al 2011). The latter are now often investigated using neutron interferometry (Colella et al 1975, Rauch and Werner 2000, Greenberger et al 2012). This technique might be adapted to detecting the new waves. The scalar field does not resemble the electromagnetic one, since its hypothetical quanta are spin-0 particles with no polarization states, rather than the spin-1 photons of light. Since the scalar field is believed to be responsible for the inertial masses of particles, it may be possible to detect it by adapting the Eotvos balance formerly used to verify the Equivalence Principle (Will 1991). Overall, it is likely that the technology necessary to probe 5D effects will more closely resemble that developed for neutrino astrophysics than the detectors of conventional optical astronomy.

The present account is preliminary. The basic theory of the two effects is discussed in Section 2, and some relevant comments are made in Section 3.

2. <u>Vacuum Waves and the Scalar Field</u>

Einstein's 4D theory of general relativity can be extended to 5D in accordance with the embedding theorem of Campbell mentioned above. The field equations for empty space in 4D in terms of the Ricci tensor are $R_{\alpha\beta} = 0$ $(\alpha, \beta = 0-3)$ and in 5D are $R_{AB} = 0$ $(A, B = 0-4)$. Here the coordinates are $x^0 = ct$ for the time, $x^{123} = x, y, z$ for ordinary 3D space and $x^4 = l$ for the extra dimension. The speed of light $c$ is kept explicit in order to bring out the physics. While space is empty of ordinary matter, it will in general have a cosmological constant $\Lambda$. In Einstein's theory, this measures the properties of a kind of vacuum fluid which is typified by an equation of state where the sum of the energy density and the pressure is zero.

Solving the aforementioned 5D field equations is a non-trivial task. But for astrophysical situations, it can be simplified somewhat by restricting the gravitational potential $g_{\alpha\beta} = g_{\alpha\beta}(x^\gamma, l)$ to the four diagonal components associated with time and the three directions in ordinary space. To these must be added the scalar potential associated with the extra dimension, which it is convenient to write as $g_{44} = \varepsilon \Phi^2$ where $\Phi = \Phi(x^\gamma, l)$ in general. Here $\varepsilon = \pm 1$ indicates whether the extra dimension is spacelike $(\varepsilon = -1)$ or timelike $(\varepsilon = +1$; there is no problem with closed timelike paths in modern 5D theory because the extra coordinate does not have the physical nature of a time). The number of unknowns to be determined by the field equations can actually be reduced further if 3D space is isotropic and homogeneous, as in cosmology, because then the spatial potentials are of the same type. Also, if the aim is to isolate waves in ordinary 3D space, it is possible to further simplify the algebra by taking the scalar potential to be smooth and flat (i.e. $\Phi = 1$; this condition has to be given up, of course, in order to study the behaviour of $\Phi$ itself). With the foregoing provisos, it is possible to look for solutions of the 5D field equations $R_{AB} = 0$ noted above. Recently, a particularly interesting kind of behavior was found (Wesson 2013) which deserves attention.

There is a solution with a constant scalar field in the extra dimension but a regular wave in ordinary 3D space. It was originally found by hand but may be quickly verified by computer. The solution is specified by the 5D interval

$$dS^2 = \frac{l^2}{L^2}\left\{c^2 dt^2 - \exp\left[\pm\frac{2i}{L}(ct+\alpha x)\right]dx^2 - \exp\left[\pm\frac{2i}{L}(ct+\beta y)\right]dy^2\right.$$

$$\left. - \exp\left[\pm\frac{2i}{L}(ct+\gamma z)\right]dz^2\right\} + dl^2 \ . \quad (1)$$

Hence the constant length $L$ is related to the cosmological constant by $\Lambda = -3/L^2$. (There is a corresponding solution with $\varepsilon = -1$ and $\Lambda = +3/L^2$, though it describes monotonic motion.) Equation (1) describes a wave in 3D, whose frequency is $f = c/L$ and whose wave-numbers in the three directions of ordinary space are $k_x = \alpha/L$, $k_y = \beta/L$, $k_z = \gamma/L$. The dimensionless constants $\alpha, \beta, \gamma$ are arbitrary, so the speed of the wave in the x-direction (say) is $c/\alpha$ and can exceed $c$. The wave of (1) is certainly not of conventional gravitational type, because apart from having a speed that can exceed $c$, it only exists if the cosmological constant is finite. That is, it depends on the energy density of the vacuum being finite. In fact, an examination of the properties of (1) shows that it closely resembles a de Broglie or matter wave, of the type found in wave/particle experiments. Among other things, it obeys de Broglie's relation $v_p v_g = c^2$ between the phase velocity and the group velocity. Usually, the former is identified with the speed of the wave and the latter is identified with the speed of the associated particle. However, a study of the dispersion relation involved shows that $v_p$ and $v_g$ have a kind of dual relationship, where the speed of light $c$ separates two modes for the behaviour of the vacuum. Further investigation is needed, but (1) shows that in principle waves can propagate at super-luminal speeds if there is a fifth dimension.

A more general case that is of interest occurs if the scalar field associated with the fifth dimension is not constant as in (1) but varies in some prescribed manner. It should be recalled that in 5D, causality may be defined by the null condition on the interval, $dS^2 = 0$. This includes the conventional 4D conditions for both photons and massive particles, $ds^2 \geq 0$. In general, therefore, causal paths obey

$$dS^2 = 0 = ds^2 + \varepsilon \Phi^2(x^\gamma, l) dl^2 \quad , \tag{2}$$

where $ds^2$ is the interval for the 4D part of any solution of the 5D field equations. Many such solutions have been found in recent years which are relevant to particles and large-scale systems. It has become evident that velocities in ordinary 3D space can exceed the speed of light if $\varepsilon = +1$ in (2) and the extra dimension is timelike. Even when $\Phi = 1$, the velocity in ordinary 3D space is then $dx/dt = [c^2 + (dl/dt)^2]^{1/2}$ and exceeds $c$. And when $\Phi \neq 1$ there are many possibilities for signalling because the scalar field may be real or complex, depending on the situation. Irrespective of the details, (2) shows that the magnitude of the element of 4D proper time is related to the scalar field by $|ds| = \Phi dl$. It is clearly possible in principle to produce a sympathetic response in $|ds|$ by varying $\Phi(x^\gamma, l)$ in a systematic way. Further, if the 4D metric tensor on which $|ds|$ depends is independent of the extra coordinate, the field equation which governs $\Phi$ reads $R_{44} = \Box \Phi = 0$ (where the box operator represents the second-order covariant divergence). This is Laplace's equation, which of course admits wave-like solutions. Thus it is possible to transmit a signal which is detectable in 4D spacetime by modulating the potential associated with the fifth dimension.

The physics associated with Equations (1) and (2) is striking. But those relations presumably refer to localized regions of the universe, and the question arises of whether the 5D theory admits solutions which are in agreement with global observations. That is, are there solutions of the 5D field equations $R_{AB} = 0$ which have the appropriate amounts of visible matter, dark matter and dark energy, as well as the correct dynamical parameters including the age? The answer to this question is affirmative, as will now be seen.

The standard class of 5D cosmologies was worked out first by Ponce de Leon (1988). He obtained 5D analogs of the 4D Friedmann-Robertson-Walker models by deriving solutions of the 5D field equations that reduced to solutions of the 4D Einstein equations on the hypersurfaces $x^4 = l$ = constants. Ponce de Leon's solutions included a 5D version of the de Sitter solution, with a positive cosmological constant as required for inflation in the early universe. He also found a group of solutions defined by a dimensionless parameter $\alpha$ and the ordinary matter of the late universe. In somewhat different notation from the original, the latter group of models has line element

$$dS^2 = l^2 dt^2 - t^{2/\alpha} l^{2/(1-\alpha)} \left( dr^2 + r^2 d\Omega^2 \right) - \alpha^2 (1-\alpha)^{-2} t^2 dl^2 \quad . \tag{3.1}$$

This has expanding, flat space sections measured in spherical polar coordinates $(r, \theta, \phi; \ d\Omega^2 \equiv d\theta^2 + \sin^2\theta d\phi^2)$, and an extra dimension which grows with coordinate time ($t$). In terms of a more physical time coordinate $(T \equiv lt)$ and the assignable dimensionless constant $(\alpha)$ the density and pressure of the matter are given by

$$8\pi\rho = 3/\alpha^2 T^2, \quad 8\pi p = (2\alpha - 3)/\alpha^2 T^2 \quad . \tag{3.2}$$

These describe radiation for $\alpha = 2$ and dust for $\alpha = 3/2$.

For 5D models with more exotic equations of state, it is necessary to employ a metric which more closely resembles (1) above, though now describing an expanding fluid with a large vacuum component. The framework for models of this type was laid out by Liu and Mashhoon (1995). In fact, there are several different models available which agree with observational data for the present universe but allow of unusual behavior for the early universe, including one where the big bang is replaced by a big bounce (Liu and Wesson 2001). However, the most important thing is that solutions of the 5D field equations are known that have a mixture of vacuum and matter fields which evolve in a manner that is in agreement with the latest observations (Zhang et al 2006 and references therein). Thus the wave and scalar field described by (1) and (2) are part of a consistent 5D approach to cosmology.

The extension from 4D to 5D also means that some long-standing problems in cosmology are avoided. For example, the horizon problem consists in the fact that the temperature of the 3K microwave background is currently uniform over the sky to high accuracy, even though according to standard models of general relativity the photons were emitted from regions that early on were causally disconnected. A similar argument has been made for the values of the particle parameters in high-redshift objects such as QSOs (Tubbs and Wolfe 1980). In dimensionally-extended cosmology, such contradictions do not arise. This because the 4D definitions of causality, as $ds^2 = 0$ for photons and $ds^2 > 0$ for massive particles, are replaced by the 5D definition $dS^2 = 0$ for *all* particles. That is, everything in the 5D universe is causally connected to everything else.

3. <u>Conclusion and Discussion</u>

If there are more than four dimensions, as needed for the unification of classical and quantum physics, it is feasible that signals can propagate at greater than lightspeed. This is a general property of a 5D metric with the appropriate signature, as employed in Space-Time-Matter theory and Membrane theory. Specifically, there is an exact solution of the 5D field equations (1) which describes a wave travelling through ordinary 3D space at more than the velocity of light. Alternatively, the modulation of the scalar field (potential) which forms the fifth dimension can affect the spacetime interval via (2) in ways that effectively travel at greater than lightspeed.

Detectors of a type able to measure such super-luminal signals have yet to be developed. However, the 5D vacuum waves of (1) resemble 4D de Broglie waves, and might be measured by a device like a neutron interferometer. While modulating the scalar field of the fifth dimen-sion would by (2) cause a sympathetic response in the 4D interval or proper time, which might be measured by an inertial balance or an atomic clock.

The preceding comments are made with the development in mind of detectors in an Earth-based laboratory. However, it will doubtless have occurred to the astute reader that there could be other civilizations out there who are already using super-luminal signals for interstellar communication. Indeed, the thought occurs that the Search for Extraterrestrial Intelligence using Earth-based telescopes has not detected any messages in the 4D electromagnetic mode because the aliens are using a different (5D) mode.

Communication with extraterrestrial civilizations using electromagnetic waves must inevitably be compromised to some extent by the finite speed of light. If the universe really does have five (or more) dimensions, exchanging signals via the fifth dimension may be the faster and preferred method.

Acknowledgements

Thanks for comments on 5D waves and neutron interferometry go to J.M. Overduin and S.A. Werner. Background material on 5D physics is available online at 5Dstm.org. Version 2 of this ArXiv paper includes material from Version 1 (ArXiv 1401.2883) and material from a paper on the same subject (Physics International $\underline{5}$, 5, 2014).